\newif\ifabstract
\abstracttrue
\abstractfalse 
\newif\iffull
\ifabstract \fullfalse \else \fulltrue \fi

\documentclass[11pt]{article}

\advance \oddsidemargin by -0.8in
\newcommand{\myparskip}{3pt}
\parskip \myparskip

\usepackage[T1]{fontenc}
\usepackage{color,colortbl}
\usepackage{amsmath}
\usepackage{cleveref}

\usepackage{multirow}   
\usepackage{hhline}
\usepackage{array}
\usepackage{pifont}
\usepackage{enumerate}

\usepackage{graphicx,epsfig,amsfonts,bbm,epstopdf,tabularx}
\usepackage{subfigure}

\usepackage{latexsym}
\usepackage{url}

\usepackage{xcolor}

\usepackage{algorithm,algpseudocode}
\usepackage{algorithmicx}
\usepackage{stmaryrd}
\usepackage{bm}
\usepackage{multirow}
\usepackage{graphicx}
\usepackage{arydshln}
\usepackage{mathtools}
\usepackage{tabulary}
\usepackage{booktabs}
\usepackage{subfigure}
\usepackage{xspace}

\newtheorem{ass}{\textbf{Assumption}}

\newtheorem{thm}{\textbf{Theorem}}

\newcommand{\CR}{\texttt{CR}\xspace}

\newcommand{\opt}{\texttt{OPT}\xspace}
\newcommand{\alg}{\texttt{ALG}\xspace}
\newcommand{\ota}{\texttt{OTA}\xspace}

\newcommand{\oa}{\texttt{OA}\xspace}

\newcommand{\ok}{\texttt{OKP}\xspace}

\newcommand{\okd}{\texttt{OKD}\xspace}

\newcommand{\cali}{\mathcal{I}}
\newcommand{\calt}{\mathcal{T}}

\newcommand{\bx}{\boldsymbol{x}}
\newcommand{\ud}{\overline{D}}
\newcommand{\ld}{\underline{D}}

\definecolor{DarkGreen}{RGB}{0, 100, 0}

\usepackage{ifthen}
\newboolean{showcomments}
\setboolean{showcomments}{true}

\newcommand{\rev}[1]{\ifthenelse{\boolean{true}}
	{\textcolor{red}{  #1 }}{}}
\newcommand{\addcite}[0]{\ifthenelse{\boolean{showcomments}}
	{\textcolor{blue}{~(add cite(s)) } }{}}

\begin{document}
	
	\title{The Online Knapsack Problem with Departures}
	
	
	
	
	
	\def\thefootnote{\#}\footnotetext{These authors contribute equally to this work.}\def\thefootnote{\arabic{footnote}}
	
	\author{Bo~Sun$^\#$\thanks{The Chinese University of Hong Kong. Email: {\tt bsun@cse.cuhk.edu.hk}.}\and 
		Lin~Yang$^\#$\thanks{Nanjing University. Email: {\tt linyang@nju.edu.cn}.} \and
		Mohammad~Hajiesmaili\thanks{University of Massachusetts Amherst. Email: {\tt hajiesmaili@cs.umass.edu}.} \and
		Adam~Wierman\thanks{California Institute of Technology. Email: {\tt adamw@caltech.edu}.} \and
		John~C.S.~Lui \thanks{The Chinese University of Hong Kong. Email: {\tt cslui@cse.cuhk.edu.hk}.} \and
		Don~Towsley \thanks{University of Massachusetts Amherst. Email: {\tt towsley@cs.umass.edu}.} \and
		Danny~H.K.~Tsang\thanks{The Hong Kong University of Science and Technology and The Hong Kong University of Science and Technology (Guangzhou). Email: {\tt eetsang@ust.hk}.}
	}

	\begin{titlepage}
		\maketitle

		\thispagestyle{empty}
		
		\begin{abstract}
			The online knapsack problem is a classic online resource allocation problem in networking and operations research. Its basic version studies how to pack online arriving items of different sizes and values into a capacity-limited knapsack. In this paper, we study a general version that includes \textit{item departures}, while also considering \textit{multiple knapsacks} and \textit{multi-dimensional item sizes}. We design a threshold-based online algorithm and prove that the algorithm can achieve order-optimal competitive ratios. Beyond worst-case performance guarantees, we also aim to achieve near-optimal average performance under typical instances. Towards this goal, we propose a data-driven online algorithm that learns within a policy-class that guarantees a worst-case performance bound. In trace-driven experiments, we show that our data-driven algorithm outperforms other benchmark algorithms in an application of online knapsack to job scheduling for cloud computing.
		\end{abstract}
	\end{titlepage}

	\maketitle
	
	\section{Problem Statement}
	We study the classic the online knapsack problem (\ok). 
	In its most basic version, there is only one knapsack, and each item is characterized by its value and one-dimensional (scalar) size.
	The problem is to irrevocably decide whether to admit each item upon its arrival with the goal of maximizing the total values of admitted items while respecting the capacity of the knapsack.
	The sequence of items can only be revealed one-by-one and may even be adversarial.
	In this paper, we focus on a novel generalization that includes item departures, while also considering multiple knapsacks and multi-dimensional item sizes.

	Consider $K$ knapsacks in a slotted time horizon $[T] = \{1,\dots,T\}$, where each knapsack $k\in[K]$ has capacity $C_k\in\mathbb{R}^+$. A total of $N$ items arrive sequentially and each item $n$ is characterized by its item information $\cali_n = \{a_n, \{w_{nk},v_{nk},\calt_{nk}\}_{k\in[K]}\}$, where $a_n$ is the arrival time, and for each knapsack $k$, $w_{nk}$ and $v_{nk}$ are the size and value, and $\calt_{nk}:=\{s_{nk},\dots,s_{nk}+d_{nk}-1\}$ is the set of time slots that item $n$ requests to stay in knapsack $k$ from its starting time $s_{nk}$ to its departure time $s_{nk}+d_{nk}-1$. 
	{The set $\calt_{nk}$ contains $d_{nk}$ consecutive time slots and we call $d_{nk}$ the duration of the item. }

	Upon arrival of item $n$, a decision maker observes its item information $\cali_n$ and determines (i) whether to admit this item, and (ii) which knapsack this item should be assigned to if it is admitted.
	Let $\bx_n = \{x_{nk}\}_{k\in[K]}$ denote the decision variable, where $x_{nk}\in \{0,1\}$ indicates whether to admit item $n$ to knapsack $k$ and $\sum_{k\in[K]}x_{nk} = 0$ represents declining the item.
	The goal is then to design an online algorithm to causally determine $\bx_n$ based on the item information up to $n$, i.e., $\{\cali_{n'}\}_{n'\le n}$, that maximizes the total value of all admitted items while ensuring the capacities of all knapsacks not to be violated over the time horizon.
	
	Let $\cali:=\{\cali_n\}_{n\in[N]}$ denote an instance of the online knapsack with departures (\okd). 
	Given $\cali$, the offline problem is shown as
	\begin{subequations}
		\label{p:offline}
		\begin{align}
			\max_{x_{nk}} \quad&\sum\nolimits_{n\in[N]}\sum\nolimits_{k\in[K]} v_{nk} x_{nk},\\
			{\rm s.t.}\quad& \sum\nolimits_{n\in[N]: t\in\calt_{nk}} w_{nk} x_{nk} \le C_k, \forall k\in[K], t\in [T],\\
			& \sum\nolimits_{k\in[K]} x_{nk} \le 1, \forall n\in[N],\\
			&x_{nk} \in\{0,1\}, \forall n\in [N], k\in[K].
		\end{align}
	\end{subequations}
	Let $\opt(\cali)$ and $\alg(\cali)$ denote the values obtained by the offline problem~\eqref{p:offline} and an online algorithm under the instance $\cali$. The competitive ratio of the online algorithm is defined as the worst-case performance ratio of the offline and online algorithms, i.e., ${\CR = \max_{\cali} \opt(\cali)/\alg(\cali)}$. Our goal is to design an online algorithm that can achieve the smallest competitive ratio.
	
	We make following technical assumptions.
	
	\begin{ass}[Value density fluctuation]\label{ass:value-density}
		The value density of each item $n$ in knapsack $k$ is bounded, i.e., ${v_{nk}}/({w_{nk} d_{nk}}) \in [1,\theta_k], \forall n\in[N]$ and value density ratio is defined as $\theta_k$.
	\end{ass}
	\begin{ass}[Duration fluctuation]\label{ass:duration}
		The duration of each item $n$ in knapsack $k$ is bounded, i.e., $d_{nk} \in [\ld_k, \ud_k]$, $\forall n\in[N]$ and duration ratio is defined as $\alpha_k = \ud_k/\ld_k$.
	\end{ass}
	\begin{ass}[Upper bound of item size]\label{ass:size}
		The size of each item $n$ is upper bounded, i.e., ${w_{nk} \le \varepsilon_k \le C_k, \forall k\in[K], n\in[N]}$.
	\end{ass}
	
	\section{Algorithms \& Main Results}
	\begin{algorithm}[H]
		\caption{Online Algorithms for Online Multiple Knapsacks with Departures (\oa-\okd)}
		\label{alg:ota}
		\begin{algorithmic}[1]
			\State \textbf{input:} threshold function $\phi= \{\phi_k\}_{k\in[K]}$, knapsack capacities $\{C_k\}_{k\in[K]}$; 
			\State \textbf{output:} admission and assignment decision $\bx_n = \{x_{nk}\}_{k\in[K]}$;
			\State \textbf{initialization:} initial knapsack utilization $z_{kt}^{(0)} = 0, \forall k, t$;
			\For{each item $n$ with item information $\cali_n$}
			\For{each knapsack $k\in[K]$}
			\label{algline:check-admit1}
			\State call Algorithm~\ref{alg:oac} to check admissibility 
			$\hat{x}_{nk} = \emph{\ota}(\cali_n,\phi_k,\{z_{kt}^{(n-1)}\}_{t\in\calt_{nk}},C_k)$;
			\label{algline:call-alg2}
			\EndFor
			\label{algline:check-admit2}
			\If{$\sum_{k\in[K]}\hat{x}_{nk} > 0$}
			\label{algline:assign-admit1}
			\State admit item $n$ and assign it to knapsack $k' = \arg\max_{k\in[K]:\hat{x}_{nk}=1} v_{nk}$; \label{algline:admit}
			\State set $x_{nk'} = 1$ and $x_{nk} = 0, \forall k\in[K]\setminus \{k'\}$;
			\Else 
			\State decline item $n$ and set $x_{nk} = 0, \forall k\in[K]$; \label{algline:decline}
			\EndIf
			\label{algline:assign-admit2}
			\State update knapsack utilization $z_{kt}^{(n)} = z_{kt}^{(n-1)} + w_{nk} x_{nk}, \forall k\in[K], t\in\calt_{nk}$.
			\label{algline:update}
			\EndFor
		\end{algorithmic}
	\end{algorithm}
	\begin{algorithm}[t]
		\caption{Online Threshold-based Algorithm for Admission Control (\ota)
		}
		\label{alg:oac}
		\begin{algorithmic}[1]
			\State \textbf{input:} item information $\{v,w,\calt\}$, threshold function $\phi$, utilization $\{z_{t}\}_{t\in\calt}$, capacity $C$;
			\State \textbf{output:} admission decision $\hat{x}$;
			\State determine a threshold value $\Phi = \sum_{t\in\calt} w \phi(z_t)$;
			\label{algline:threshold-value}
			\If{$v \ge \Phi$ and $z_t+w \le C, \forall t\in\calt$}
			\label{algline:ota-check}
			\State item is admissible and set $\hat{x} = 1$;
			\Else
			\State item is inadmissible and set $\hat{x} = 0$.
			\EndIf
		\end{algorithmic}
	\end{algorithm}
	
	We propose a simple yet effective online algorithm (\oa-\okd) to solve \okd in Algorithm~\ref{alg:ota}.
	It consists of two parts: decomposing the multiple knapsack problem into the admissibility check of each individual knapsack and admission control of each individual knapsack via an online threshold-based algorithm (\ota). The key step is the admission control of items to each knapsack  by calling the \ota subroutine in Algorithm~\ref{alg:oac}.
	To check admissibility, \ota defines a threshold value (line~\ref{algline:threshold-value}) as $\Phi = \sum\nolimits_{t\in\calt} w \phi\left(z_t\right)$, where $\phi(z_t)$ can be interpreted as the marginal cost of the unit item if it stays in the knapsack in slot $t$, and is a function of the real-time knapsack utilization $z_t$. 
	Since $\ota_\phi$ is fully parameterized by $\phi$, the key design question is how to determine the threshold function $\phi$ such that \oa-\okd is competitive with the offline optimum.
	
	By carefully designing the threshold function, our main results can be summarized as follows.
	\begin{thm}\label{thm:up-omkd}
		Under Assumptions~\ref{ass:value-density}-\ref{ass:size}, there exists a parameter $\gamma_k = O(\ln(\alpha_k\theta_k)), \forall k\in[K]$, if the item size is upper bounded by $\varepsilon_k \le {C_k\ln2}/{\gamma_k}, \forall k\in[K]$, and the threshold function is given by $\phi^\gamma := \{\phi_k^{\gamma_k}\}_{k\in[K]}$, where
		\begin{align}\label{eq:threshold}
			\phi_k^{\gamma_k}(z) = \exp\left(z \gamma_k/C_k \right) - 1, z\in[0,C_k], \forall k\in[K],
		\end{align}
		then the competitive ratio of the algorithm $\oa(\phi^{\gamma})$ is $O(\ln(\alpha\theta))$, where $\theta = \max_{k\in[K]} \theta_k$ and $\alpha = \max_{k\in[K]} \alpha_k$. 
	\end{thm}
	\begin{thm}\label{thm:lb}
		There is no online algorithm that can achieve a competitive ratio smaller than $\Omega(\ln(\alpha\theta))$ for the online multiple one-dimensional knapsacks with departures.
	\end{thm}
	Combining the upper bound result in Theorem~\ref{thm:up-omkd} and the lower bound result in Theorem~\ref{thm:lb}, we conclude that our proposed $\oa(\phi^{\gamma})$ achieves an order-optimal competitive ratio for \okd.
	In the full paper~\cite{sun2022online}, this proposed algorithm can be further extended to consider the multi-dimensional item size and also achieve the order-optimal competitive ratio. In addition, we additionally design a data-driven online algorithm that can adaptively select the parameter $\gamma$ such that the overall algorithm can work well on practical instances and, in the meanwhile, provide the worst-case guarantees. 

\bibliographystyle{abbrv}
\bibliography{references}
	
\end{document}